\documentclass[
 amsmath,amssymb,
 aps,
prb,
]{revtex4-2}
\usepackage{graphicx} 
\usepackage{CJKutf8}
\usepackage{mathtools}
\usepackage{placeins}
\usepackage{comment}
\usepackage[normalem]{ulem}
\usepackage[dvipsnames,usenames]{xcolor}

\newcommand{\LANL}{Los Alamos National Laboratory, Los Alamos, New Mexico, 87545, USA}
\newcommand{\nuebar}{\overline{\nu}_e}

\newcommand{\dd}{\textrm{d}}

\begin{document}

\title{Novel Application of Neutrinos to Evaluate U.S. Nuclear Weapons Performance}
\author{J.R. Distel}
\affiliation{\LANL}
\author{E.C. Dunton}
\affiliation{\LANL}
\author{J.M. Durham}
\affiliation{\LANL}
\author{A.C. Hayes}
\affiliation{\LANL}
\author{W.C. Louis}
\affiliation{\LANL}
\author{J.D. Martin}
\affiliation{\LANL}
\author{G.W. Misch}
\affiliation{\LANL}
\author{M.R. Mumpower}
\affiliation{\LANL}
\author{Z. Tang}
\affiliation{\LANL}
\author{R.T. Thornton}
\affiliation{\LANL}
\author{B.T. Turner}
\affiliation{\LANL}
\author{R.G. Van de Water}
\affiliation{\LANL}
\email{vdwater@lanl.gov}
\author{W.S. Wilburn}
\affiliation{\LANL}

\begin{abstract}
There is a growing realization that neutrinos can be used as a diagnostic tool to better understand the inner workings of a nuclear weapon.  
Robust estimates demonstrate that an Inverse Beta Decay (IBD) neutrino scintillation detector built at the Nevada Test Site of 1000-ton active target mass at a standoff distance of 500\,m would detect thousands of antineutrino events per nuclear test. 
This would provide less than 4\% statistical error on measured antineutrino rate and 5\% error on antineutrino energy.  Extrapolating this to an error on the test device explosive yield requires knowledge from evaluated nuclear databases, non-equilibrium fission rates, and assumptions on internal neutron fluxes. Initial calculations demonstrate that the total number of neutrinos emitted per fission in the first $10^{3}$s after a short pulse of $^{239}$Pu fission is about a factor of two less than that from Pu fissioning under steady state conditions.  
As well, there are significant energy spectral differences as a function of time after the pulse that must be considered. These and other model dependencies will be discussed in the paper. In the absence of nuclear weapons testing, many of the technical and theoretical challenges of a full nuclear test could be mitigated with a low cost smaller scale 20 ton fiducial mass IBD demonstration detector placed near a pulsed reactor.  Potential reactors include the Texas A\&M University (TAMU) TRIGA 1GW-10\,millisecond pulsed facility or the Sandia Annular Core Research Reactor. 
The short duty cycle and repeatability of pulses would provide critical real environment testing and measurements which would be valuable for planning  a possible real test shot in the future. 
Furthermore, the antineutrino rate as a function of time data would provide unique constraints on fission databases and model assumptions. Finally, there are impactful science drivers such as sensitive searches for $\sim$1 eV$^2$ sterile neutrinos and $\sim$MeV scale axions.  
\end{abstract}
\maketitle

\section{Introduction}

In this paper, we discuss the feasibility of diagnosing a potential U.S. nuclear weapons test using a neutrino detector. This idea was first proposed in 1953 by Reines and Cowan~\cite{Reines1953, LANL1997} to observe the then hypothesized neutrino. Since then, the neutrino was discovered \cite{Cowan1956} and over the decades the field of neutrino research has matured and many properties of the neutrino have been measured such as interaction cross sections, masses, abundances, etc~\cite{ParticleDataGroup:2020ssz}.
The neutrino has been observed from the sun, supernova, nuclear reactors, accelerators, and even the Earth's core.   Interestingly, one of the most prolific sources of neutrinos, a man-made nuclear explosion, has yet to be detected.

\begin{figure}[h!]
\centering
\includegraphics[scale=0.24]{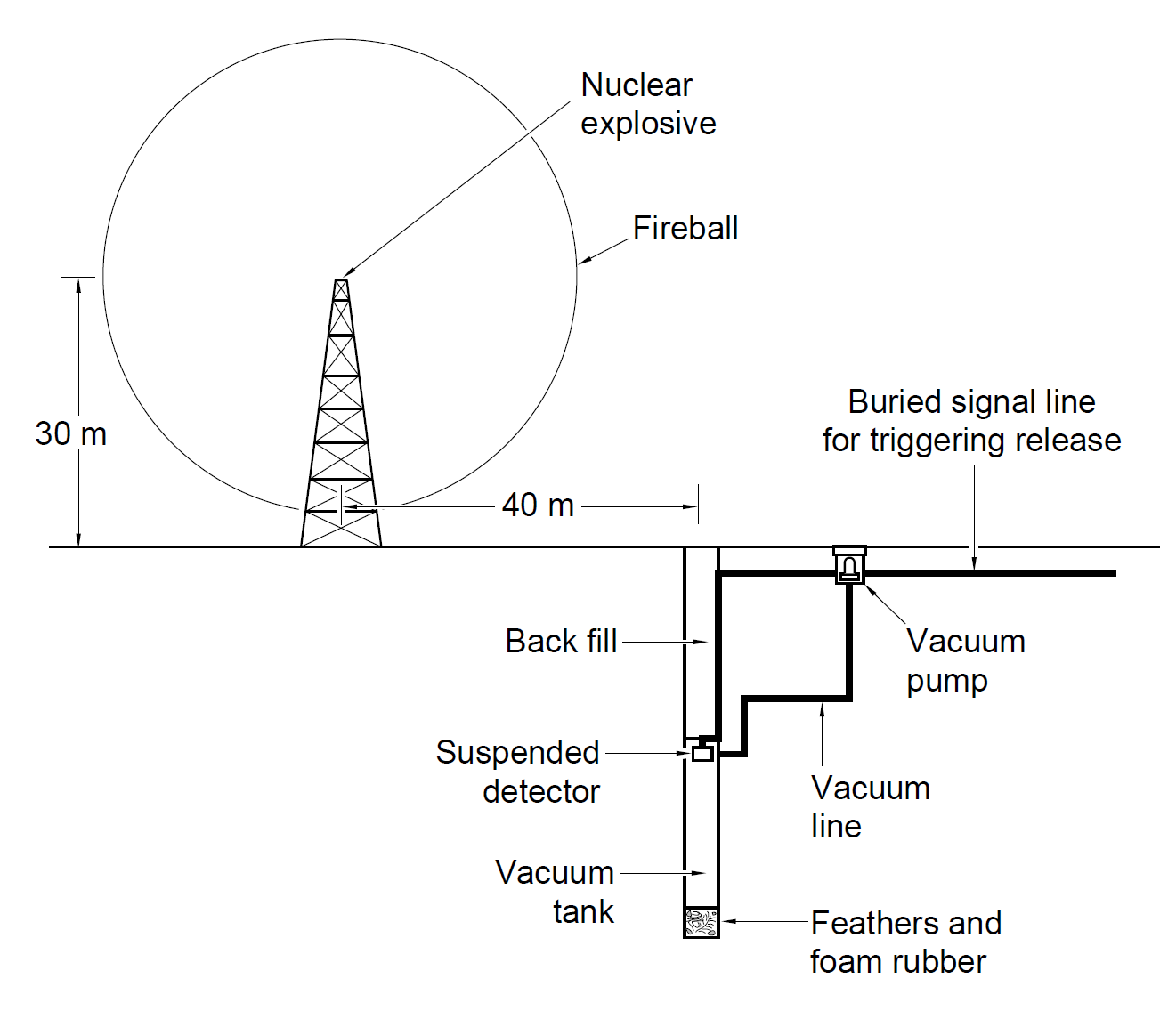}
\caption{Original proposed experiment by Reines and Cowan for the detection of the elusive neutrino. The proposed experiment is approximately 90 meters away from the nuclear explosion, necessitating a complex drop mechanism to prevent destruction of the detector before detection of the neutrinos can be achieved. Reprinted with permission from Ref.~\cite{LANL1997} Los Alamos, "The Reines-Cowan Experiments: Detecting the Poltergeist", Los Alamos Science, 25:3, 1997. Los Alamos National Laboratory, operated by Triad, LLC.}
\label{fig:reinescowan}
\end{figure}

Frederick Reines and Clyde Cowan, Jr. 
were the first physicists to understand the utility of a nuclear weapon test to produce a prolific source of neutrinos in a relatively short time window and reasonable stand off distance, as shown in Fig.\ref{fig:reinescowan}.  
Such a concept was proposed at LANL in the 1950’s~\cite{LANL1997, Abbott2021} but was finally rejected for a more controlled experiment at a nuclear reactor \cite{Reines1953,Cowan1956} – this latter method went on to win the 1995 Nobel prize in physics for the detection of the neutrino. 
What they maybe did not realize at the time is that continual improvements in neutrino detector technology over the decades could one day have many applications to LANL's main mission more than 70 years later. 

The application that we will be exploring in this paper is the use of a $\sim$1000-ton scale Inverse Beta Decay (IBD) neutrino detector to measure the nuclear weapon performance at a safe stand off distance.  Situating the  detector at  a  central accessible point allows it to be re-used for  multiple shots. A 500\,m standoff distance is sufficient to insulate the detector  from shock waves but close enough  to potentially detect thousands of antineutrinos from a test.  The detector should be at least 10\,m underground to shield from cosmic ray backgrounds. 
First, we perform robust rate calculations for antineutrinos from a hypothetical nuclear yield to indicate if the concept is feasible.  
We estimate the antineutrino flux and IBD event rates from a weapon test as a function of energy and time by assuming a single prompt fission event which generates fragments with a known independent yield distribution.  These products decay to a distribution of stable (or sufficiently long lived) isotopes and the resulting antineutrino spectrum is folded with known antineutrino cross sections to produce rates for IBD interactions.
A discussion of systematic errors associated with extrapolating the measured neutrino rates to nuclear detonation yield will be presented.

The proven detection method for neutrinos from power reactors (U or Pu fission) is a large mass liquid scintillator detector ($>$100-tons) that measures the IBD channel via $ \nuebar + p \rightarrow n + e^+ $~\cite{Qian:2018wid}.  The detection of the prompt positron followed by gamma rays from neutron capture provides a very powerful signal with excellent background rejection.  The reconstructed positron interaction preserves the original antineutrino energy and the event time to the nanosecond level.
Thus, the antineutrino flux rate, energy, and time distribution can be accurately reconstructed.  Initial straight forward rate estimates demonstrate that a large \>1000-ton-scale detector 500\,m from the nuclear test device will have significant event rates per kTe of nuclear yield~\footnote{Throughout this work we use the convention kTe for kilo-tons of  explosive yield, while for neutrino detector mass we use ton.}.  We expect that antineutrino rates can be measured with an uncertainty of four percent or better and energy at the $\sim$5\% level. 
These robust detectors might be the key to detecting neutrinos with sufficient rate, energy, and time resolution to determine weapon yield and other parameters of interest to the U.S. weapons design community.

Finally, we discuss plans to test a small 25-ton IBD detector at a pulsed reactor facility.  The aims are three fold: 1) to test experimental issues operating an IBD detector at a pulsed fission source, 2) to measure, for the first time, the antineutrino time and energy distributions from a pulsed source, and 3) to evaluate yield systematic errors from a pulsed non-equilibrium fission source using measured neutrino rates.  This will be followed by discussing possible fundamental and applied physics that can be accomplished by such a prototype detector at a pulsed reactor.

\section{Estimated detected neutrino event rates from a potential nuclear weapons test}
\label{sec:Estimate}
The fission processes produce neutrinos throughout the fission decay chain, which summed up total about $\sim$2.5 antineutrinos per U or Pu fission decay in the first $10^{3}$ seconds after a fast fission pulse as described in Section~\ref{nurates}.  For a steady state nuclear reactor this number is about $\sim$5 antineutrinos per fission (see table~\ref{tab:pu_nu_norms_time}).  

Detailed flux calculations are readily available and show antineutrino spectra up to about 10\,MeV~\cite{Huber:2011wv}.
Once produced, the antineutrinos are emitted isotropically and will pass through the ground.  However, with a sizable scintillation detector filled with hydrogen (such as mineral oil - CH$_2$), some of the antineutrinos can interact through the inverse beta decay mechanism, producing a time correlated positron and neutron in  the final state that can be detected.   
Due to the threshold of 1.806\,MeV for the reaction, the maximum energy the positron can deposit is $\sim$8\,MeV in the form of scintillation light in the detector medium.  If the detector is doped with Gd,  then the final state  neutron thermalizes quickly and captures on the Gd within 10's of micro-seconds producing up to $\sim$8\,MeV of  gamma-rays~\cite{Marti:2019dof}.  The time coincidence of  the  positron and neutron  capture signals are unique and provide powerful rejection of random backgrounds.

Decades of development of neutrino detectors to measure neutrino properties from nuclear reactors has produced a proven and robust design to detect antineutrino IBD interactions~\cite{DayaBay:2015kir,DoubleChooz:2022ukr}. The basic detector design is a large volume (10 to 10,000 tons) of optically transparent scintillator material surrounded by a large number of Photo-Multiplier Tubes (PMT's)  that detect scintillation light from particle interactions.  
Fast digitizer electronics are used to read out the PMT's which allows reconstruction of antineutrino event rates, energy, position, and time of arrival.   The next subsection presents a simple antineutrino IBD rate calculation for a mineral oil based detector instrumented with hundreds of PMT's.  We start with back of the envelope calculations then move up to more complete calculations folding flux and cross sections with efficiency, ensuring consistency along the way.

\subsubsection{Estimating the measurable antineutrino signal}


We can estimate the number of antineutrino events that a typical nuclear fission explosion would generate per kiloton of TNT explosive equivalent nuclear energy (kTe) released in an IBD detector of 1000-ton typical fiducial mass at a 500m typical standoff distance.  First, the IBD event rate is given by $N_{\nu}= \phi_{\nu} \sigma N_{\textrm{free} H}$ where $N_{\textrm{free} H}$ is the number of free hydrogen targets in the detector.  The typical flux averaged cross section $\sigma$ is roughly $0.5 \times 10^{-42}$ cm$^{2}$ for a typical U or Pu fission decay neutrino flux~\cite{Anna}.  The antineutrino flux $\phi_{\bar{\nu}}$ from a burst is estimated easily from energy considerations where 1\,kTe
of TNT $= 4.2\times10^{12}$\,J $= 2.6\times10^{25}$\,MeV of energy. With about 200\,MeV/fission \cite{krane1991introductory}, this corresponds to $1.3\times10^{23}$ fissions. Including a factor of 2.5 for the average number of antineutrinos per fission emitted in the first $10^{3}$ s after a pulse of fissions induced by ``fast" neutrons (e.g. neutrons emitted during fission carrying energies between 0.4-2 MeV), as described in Section~\ref{nurates}, this gives about half a mole of antineutrinos per 1\,kTe.  Using this, the flux at $500$ m from the explosion is estimated to be $\phi_{\bar{\nu}} = 0.95 \times 10^{13}$ $\bar{\nu}$ cm$^{-2}$. A distance of 500\,m was considered a reasonable standoff distance for ensuring detector integrity, assuming that both the blast and detector are buried underground. The number of free $H$ in one metric ton of neutrino detector mineral oil is $8.6\times10^{28}$ atoms.  Putting these three numbers together gives an IBD neutrino rate of 0.41\,events/kTe/ton at 500\,m.  This is assuming 100\% efficiency.  However, the reaction threshold for neutrino IBD reactions results in a  50\% cross section efficiency above this energy.  The detector reconstruction efficiency is estimated to be about 80\% (see subsection \ref{detectorsim}), which results in a basic efficiency corrected neutrino rate of 0.16\,events/kTe/ton at 500\,m.  This translates into thousands of detected neutrino events for a 1000-ton neutrino detector at a distance of 500\,m from a plausible nuclear test~\cite{osti_1340968}.
As well as rate, an IBD detector would be able to determine neutrino energy better than 5\% resolution and event arrival time to nanosecond level.


\begin{figure*}[b] 
 \centering
\includegraphics[width=6.0in]{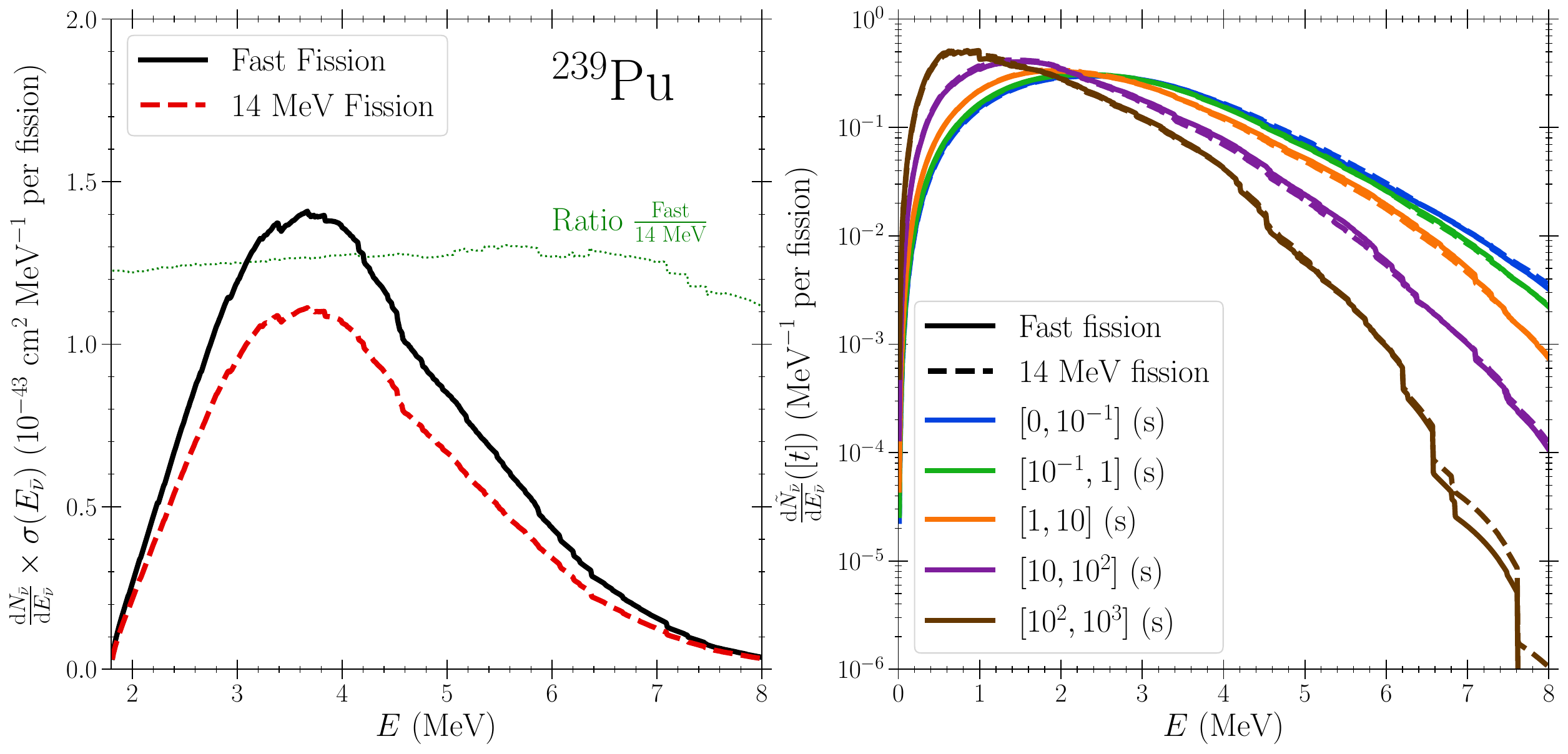}
 \caption{(Left) The time integrated and IBD cross section weighted antineutrino spectra from neutron-induced fission pulse of $^{239}$Pu for fast (black) and 14 MeV (red) neutrons. (Right) The antineutrino spectra for fissions induced by fast (solid lines) and 14 MeV (dashed lines) neutrons, integrated over five windows in time: 0-0.1 seconds (blue), 0.1-1.0 seconds (green) 
 1.0-10.0 seconds (orange), 10.0-100.0 seconds (purple), and 100.0-1000.0 seconds (grey).  The spectra have been normalized such that the integral of the shown curve over energy is equal to 1.  The extracted normalization is tabulated in table~\ref{tab:pu_nu_norms_time}.}
  \label{fig:comp-fast-14}
\end{figure*}

\section{Measurements, Extraction of Nuclear Test Yield, Systematic Errors}
\label{nurates}

In a reactor, thermal neutrons induce fissions of the actinide fuel resulting in a nearly steady state (herein abreviated as ``S.S.") emission of antineutrinos.  The weighted product of this antineutrino spectrum with the IBD cross section is known to an accuracy of approximately 3\% at its peak value, but the uncertainties are energy dependent. At the highest energies (above $\approx$8 MeV) the uncertainty is approximately 20\% \cite{an2017improved}.
The typical energies of neutrons which induce fission in a nuclear explosion are considerably higher than in a reactor environment.  In a weapon environment, the fissioning neutrons are dominantly a combination of ``fast" (e.g. those emitted from a fissioning nucleus and carrying an energy of $\sim$0.4-2.0 MeV) and 14 MeV in energy.

Extrapolating from measured reactor thermal spectra to those expected for weapons-relevant neutron energies, with reasonable accuracy, is made possible by the very detailed systematic studies of the energy-dependence of both the independent and cumulative fission product yields (FPY's) that have been carried out at Los Alamos and elsewhere, in a joint long-term experimental and theoretical program~\cite{okumura2022energy}.

\subsubsection{Extracting fission yields from the emitted neutrinos}
The spectra of neutrinos emitted by the decay of fission products generated by fast and 14 MeV neutrons are nearly equal in shape, but can be differentiated by their total magnitudes.  To see this, we assume that the fission event occurs in an extremely short duration at early time and then track the subsequent beta decays of all isotopes produced by the immediate fission event and the subsequent decays.  Using the time-dependent yields of the various unstable isotopes, we compute the evolution of the spectrum up to $10^3$ seconds after the initial fission.  The process by which this computation is performed is discussed further in the next subsection.  

In the left panel of Fig.~\ref{fig:comp-fast-14}, we predict the time-integrated antineutrino energy spectrum weighted by the IBD cross section.  As can be seen, the fast neutron induced fission antineutrino spectrum is a factor of about 1.2 higher in magnitude than the 14 MeV spectrum. In the right panel, we compare the spectral shapes of the fast and 14 MeV spectra. To make this comparison, we integrated over several different windows in time (indicated by color) and we then normalize each time-integrated spectrum such that its integral over energy is equal to one.  
The extracted normalization is then the total number of neutrinos emitted in that particular window of time.  The normalized spectrum (denoted $\textrm{d} \tilde{N}_{\bar{\nu}} / \textrm{d} E_{\bar{\nu}}$) is then plotted for the fast neutron induced fissions (solid lines) and for the 14 MeV neutron induced fissions (dashed lines).  We observe in this figure that the spectra are nearly identical implying that it is challenging to leverage information about the spectral shapes independently of the overall normalizations of the spectra to discriminate between the energies of the neutrons which induced the fission events.  We have tabulated the extracted normalizations and present them in Table~\ref{tab:pu_nu_norms_time}.
\begin{table*}
\centering

\begin{tabular}{|c||c|c|c|c|c|c|||c|}
\hline
  $^{239}$Pu  & 0.0-0.1 s & 0.1-1.0 s & 1.0-10.0 s & 10.0-100.0 s & 100.0-1000.0 s & Total & S.S. \\
\hline
\hline
 $N_{\bar{\nu}}$ (Fast)    & 0.020 & 0.145 & 0.572 & 0.890 & 0.864 &  2.49 & 4.34 \\
 \hline
 $N_{\bar{\nu}}$ (14 MeV)  & 0.018 & 0.129 & 0.479 & 0.697 & 0.707 &  2.03 & 3.63 \\
 \hline
 Fast / 14 MeV             & 1.11  & 1.12  & 1.19  & 1.27  & 1.22  &  1.23 & 1.19 \\
 \hline
 \hline
 \hline
 $^{235}$U  &  &  &  &  &  &  & \\
 \hline
 \hline
 $N_{\bar{\nu}}$ (Fast) & 0.032 & 0.219 & 0.774 & 1.013 & 0.856  & 2.89 & 4.80 \\
 \hline
\end{tabular} 
 \caption{Total number of antineutrinos integrated over energy per fission emitted in each window of time (columns) by $^{239}$Pu fission induced by fast neutrons (top row) and by 14 MeV neutrons (second row) and by $^{235}$U fission induced by fast neutrons (bottom row).  These quantities represent the total normalizations extracted from the curves presented in the right panel of Figure~\ref{fig:comp-fast-14}.  We also show the ratio of neutrinos induced by fast and 14 MeV fission events in $^{239}$Pu in each time window in the third row. The rightmost column labeled ``S.S."  (abbreviating ``steady state") is the equivalent value for the indicated isotope undergoing a constant-rate fissioning induced by the indicated neutron energy.}  \label{tab:pu_nu_norms_time}

\end{table*}

An IBD detector is ultimately sensitive only to neutrinos with energies above the IBD threshold of approximately 1.8 MeV.  The total number of events seen by the detector between times $t_{0}$ and $t_{1}$ is proportional to the spectrally averaged IBD cross section $\bar{\sigma}$.  Explicitly, 
\begin{equation} \label{eq:sigBar_def}
    \bar{\sigma}\left[ t_{0}, t_{1} \right] = \int_{t_{0}}^{t_{1}}  \dd t \int_{E_{\bar{\nu}}} \dd E_{\bar{\nu}} \frac{\dd^{2} N_{\bar{\nu}}(t,E_{\bar{\nu}})}{\dd E_{\bar{\nu}} \dd t} \sigma_{\textrm{IBD}}(E_{\bar{\nu}}) \, .
\end{equation}
Assuming no variation in the total neutrino flux across the volume of the detector which is situated a distance $R$ from the prompt fission event (e.g. a nuclear explosion), the total number of detectable IBD events in a window of time ($\left[ t_{0}, t_{1} \right]$) is then
\begin{equation} \label{eq:totEvents}
    N_{\textrm{IBD}} = N_{p}  N_{f}\frac{\bar{\sigma}\left[ t_{0}, t_{1} \right] }{4 \pi R^{2}} \, .
\end{equation}
Here, $N_{p}$ is the total number of protons in the detector and $N_{f}$ is the total number of fissions which happened in the pulse.  Thus, $\bar{\sigma}[t_{0},t_{1}]$ is the (time dependent) cross sectional area that the IBD detector presents to the evolving spectrum of antineutrinos in that window of time after the explosion.  
We have tabulated it for the same windows in time as table~\ref{tab:pu_nu_norms_time} and present it in table~\ref{tab:pu_specAvgCrossSec_time}. 

There are two opposing effects at work in the evolution of $\bar{\sigma}$.  
Because $\beta$-decay is a weak decay process, the $\beta$-decay of fission fragments occurs on a much slower timescale than the fission event itself and thus there are relatively fewer antineutrinos emitted at early times.  
However, the early-time antineutrinos carry a higher energy on average than those at later times.  In the time between $t_0$ and $t_1$, the average antineutrino energy is given by
\begin{equation} \label{eq:avgE_Twindow}
    \langle E_{\bar{\nu}}\rangle  [t_0,t_1] = \frac{
    \int_{t_{0}}^{t_{1}}  \dd t \int_{E_{\bar{\nu}}} \dd E_{\bar{\nu}} E_{\bar{\nu}} \frac{\dd^{2} N_{\bar{\nu}}(t,E_{\bar{\nu}})}{\dd E_{\bar{\nu}} \dd t}
    }{
    \int_{t_{0}}^{t_{1}}  \dd t \int_{E_{\bar{\nu}}} \dd E_{\bar{\nu}} \frac{\dd^{2} N_{\bar{\nu}}(t,E_{\bar{\nu}})}{\dd E_{\bar{\nu}} \dd t}
    } \, .
\end{equation} 
Because the IBD cross section above the reaction energy threshold is an increasing function of energy, 
$\bar{\sigma}$ rises to a peak value after approximately $10$ seconds and then falls as both the antineutrino emission rate and average antineutrino energy fall.  We have tabulated the falling time-averaged antineutrino average energy (eq.~\ref{eq:avgE_Twindow}) in Table~\ref{tab:pu_avgEnu_time} for the same time windows.  We have not accounted for the effect of antineutrino flavor oscillations in this analysis, as the oscillation probability depends on the distance $R$ which we take as a free parameter in this discussion.  The values presented here are therefore an upper bound on the detector response, and a more careful treatment applicable to a specific detector would need to take the disappearance of $\bar{\nu}_{e}$ through flavor oscillations into account by weighting the integrand of eq.~\ref{eq:sigBar_def} by the energy and distance dependent $\bar{\nu}_{e}$ survival probability.

\begin{table*}
\centering

\begin{tabular}{|c||c|c|c|c|c|c|||c|}
\hline
  $^{239}$Pu  & 0.0-0.1 s & 0.1-1.0 s & 1.0-10.0 s & 10.0-100.0 s & 100.0-1000.0 s & Total & S.S. \\
\hline
\hline
 $\bar{\sigma} $ (Fast)  & 0.077 & 0.515 & 1.603 & 1.375 & 0.666 &  4.236 & 4.65 \\
 \hline
 $\bar{\sigma} $  (14 MeV)  & 0.072 & 0.460 & 1.288 & 0.991 & 0.540 &  3.351 & 3.79 \\
 \hline
 \hline
 \hline
 $^{235}$U  &  &  &  &  &  &  & \\
 \hline
 \hline
 $\bar{\sigma} $ (Fast) & 0.146 & 0.911 & 2.574 & 1.838 & 0.782  & 6.25 & 6.77 \\
 \hline
\end{tabular}
 \caption{The spectrally averaged and time integrated IBD cross section in units of $10^{-43}$ cm$^{2}$ (denoted $\bar{\sigma}$) for $^{239}$Pu fissions induced by fast and 14 MeV neutrons (top two rows) and for $^{235}$U fissions induced by fast neutrons (bottom row).  These are obtained by weighting the time-integrated spectra for the time windows as shown in figure~\ref{fig:comp-fast-14} by the IBD cross section and the normalization of Table~\ref{tab:pu_nu_norms_time} and integrating over the energy. To calculate the experimentally measured number of IBD events, multiply $\bar{\sigma}$ by the $1/R^2$ area normalized number of fission decays, number of free $H$ atoms, and detection efficiency. Note that $\bar{\sigma}$ already contains the IBD 1.8\,MeV threshold efficiency. The rightmost column labeled ``S.S." (abbreviating ``steady state") is the equivalent value for the indicated isotope undergoing a constant-rate fissioning induced by the indicated neutron energy. } \label{tab:pu_specAvgCrossSec_time}

\end{table*}

\begin{table*}
\centering

\begin{tabular}{|c||c|c|c|c|c|||c|}
\hline
  $^{239}$Pu  & 0.0-0.1 s & 0.1-1.0 s & 1.0-10.0 s & 10.0-100.0 s & 100.0-1000.0 s & S.S. \\
\hline
\hline
 $\langle E_{\bar{\nu}} \rangle$ (Fast)  & 2.842 & 2.765 & 2.522 & 2.042 & 1.587 & 1.61 \\
 \hline
 $\langle E_{\bar{\nu}} \rangle$ (14 MeV)  &  2.871 & 2.773 & 2.495 & 1.988 & 1.581 & 1.59 \\
 \hline
 \hline
 \hline
 $^{235}$U  &  &  &  &  &  &  \\
 \hline
 $\langle E_{\bar{\nu}} \rangle$ (Fast) &  3.020 & 2.914 & 2.651 & 2.118 & 1.633 & 1.748 \\
 \hline
\end{tabular}
 \caption{Average energy of the emitted antineutrinos from fast (top row) and 14 MeV (bottom row) neutron induced fission of $^{239}$Pu across all energies in the specified time window in units of MeV. The rightmost column labeled ``S.S."  (abbreviating ``steady state") is the equivalent value for the indicated isotope undergoing a constant-rate fissioning induced by the indicated neutron energy.} \label{tab:pu_avgEnu_time}

\end{table*}



In extracting the number of fissions from a neutrino measurement, we first consider the scenario in which there is no independent measurement/knowledge of the 14 MeV neutron yield, except from a simulation prediction of unknown accuracy. We also assume that the sole actinide fissioning is $^{239}$Pu. Clearly any uncertainty in neutron fluxes will result in nuclear explosion yield errors when extrapolating from measured neutrino fluxes.

\subsubsection{Prompt Fission Yield Assumption}
On the relatively slow timescale for neutrino emission from an nuclear explosion, the neutrinos are produced by the equivalent of a neutron pulse inducing the fission, and the shape, magnitude and timing of the spectrum is determined by the direct population and subsequent $\beta$-decay of the fission fragments. 
The fission products (FP) are produced both directly in fission, with an {\it independent} yield (denoted $Y_i(Z,A,m)$), and from the $\beta$-decays of the more neutron rich fission products of the same mass.
In order to extract the time dependent energy spectrum of antineutrinos emitted through the $\beta$-decays of the prompt FP's, we make the assumption that the FP's were populated and distributed equally to the known FP independent yields for a given fuel isotope by an approximately zero-duration event at what we define to be $t=0$.  We then track the decays of each FP along each possible decay chain until a distribution of quasi-stable (i.e.~sufficiently long lived) isotopes and isomers are populated from the initial distribution.  We obtained each species' half-life ($t_{1/2}$) and decay branching ratios ($b$) for each decay from the RIPL-3 nuclear reaction database \cite{capote2009ripl}.  We determine the time-dependent yield of each species (denoted as simply $Y(Z,A,m)$) by solving a coupled, linear, initial value problem for which the instantaneous change in the time dependent yield of an isotope with atomic number $Z$, mass $A$, and isomer index $m$ is given by
\begin{equation} \label{eq:time_dep_yi}
    \frac{\textrm{d}}{\textrm{d}t } Y(Z,A,m) = - \Gamma(Z,A,m) Y(Z,A,m) 
        + \sum_{(Z',A',m')} b(Z',A',m') \Gamma(Z',A',m') Y(Z',A',m')  + \Gamma_{f}(t) Y_{i}(Z,A,m) \, .
\end{equation}
Here, $\Gamma(Z,A,m) \equiv \ln(2) / t_{1/2}$ is the decay rate of the species $(Z,A,m)$ which has half-life $t_{1/2}$.  The sum over $(Z',A',m')$ represents the total contribution of all possible species which {\it decay into} the left hand side (LHS) species $(Z,A,m)$, and 
the factor $b(Z',A',m')$ represents the branching ratio for which the parent species undergoes the specific decay to the 
LHS species. The sum includes branching over all possible transitions between species such as: internal conversions between isomer states, beta decay, and beta decay with emitted neutrons.

The prompt fission assumption enters the model through the initial condition (i.e. at $t=0$ the yields are given by the independent yields of the fissioning isotope) and the choice $\Gamma_{f}(t) = 0$ in eq.~\ref{eq:time_dep_yi}.
If fissioning is extended in time, then the source term must be accounted for in tracking the evolution of each species' time dependent yield. 

A fission nuclear reactor operating at steady state fissions actinide isotopes at a nearly constant rate, resulting in $\Gamma_{f}$ taking a constant value. After the reactor reaches steady state, the rate at which any individual isotope decays is equal to the rate at which it is populated both directly from fissions and from the decay of other isotopes.  For unstable isotopes the value to which a yield saturates is proportional to the ratio of the steady state fissioning rate to the species' half-life and the species \textit{cumulative} yield.  The cumulative yield is simply the sum of the independent yields of all of the isotopes in the decay chain ``above" the isotope under consideration (suitably weighted by the branching ratios along the decay chain).  The cumulative yields $^{239}$Pu and $^{235}$U are known \cite{Chadwick20112887short} and are utilized to model neutrino fluxes from reactors operating under equilibrium conditions \cite{HayesVogel}.  
Under such steady state conditions, the dynamic cross sections, number of neutrinos per fission, and average neutrino energy all become time independent and we provide them for ease of comparison in the rightmost columns labeled ``S.S." in tables \ref{tab:pu_nu_norms_time}, \ref{tab:pu_specAvgCrossSec_time}, and \ref{tab:pu_avgEnu_time}.  In the evaluation of the quantities we report, we include in the antineutrino emission processes only the $\beta$-decays of nuclei; we have not included the decay of any present unbound neutrons in the S.S. system.

\begin{figure*}[h] 
   \centering
   \includegraphics[width=6in]{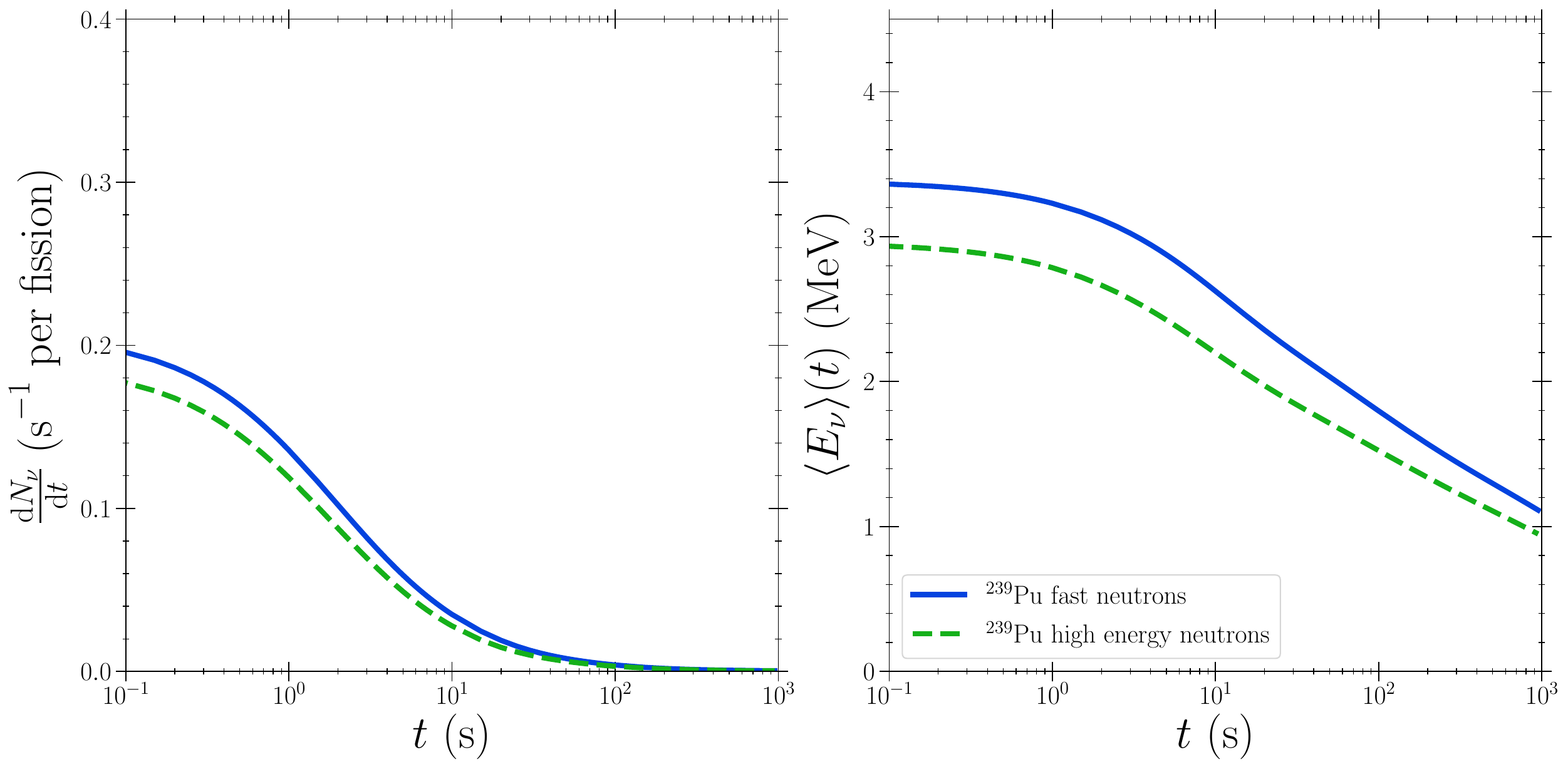} 
   \caption{The total antineutrino emission rate (left) and the instantaneous average energy (right) of neutrinos emitted by the chain of beta decays of promptly generated fission products for fast and 14 MeV induced fissions of $^{239}$Pu.  The legend applies to both panels.}
   \label{fig:NuTime}
\end{figure*}

Once the time-dependent fission product yields are known, the instantaneous total antineutrino energy spectrum (denoted $\mathcal{S}(E,t)$) is then the weighted sum over all of the $\beta$-decay spectra from the present decaying nuclei 
\footnote{We have further assumed that all considered beta decays are allowed.  A more detailed analysis including shape factor corrections to forbidden decays will be presented in a future work.}, 
and the weights are a product of the time dependent yield fractions determined above and the total $\beta$-decay branching ratio for the decay of each species.  
If each species' antineutrino energy spectrum is denoted $\mathcal{S}(E;Z,A,m)$ then the total instantaneous spectrum is 
\begin{equation}
    \mathcal{S}(E,t) = \sum_{Z,A,m} b_{\beta}(Z,A,m) \mathcal{S}(E;Z,A,m)Y(Z,A,m;t) .
\end{equation}
 The instantaneous average neutrino energy is then
\begin{equation}
    \langle E_{\bar{\nu}} (t) \rangle = \frac{\int \mathcal{S}(E,t) E \dd E}{\int \mathcal{S}(E,t) \dd E} .
\end{equation}
The total number of antineutrinos emitted ($N_{\bar{\nu}}$) per unit time (e.g. the total antineutrino emission rate) is the sum of the $\beta$-decay rates of the individual species such that
\begin{equation}
    \frac{\dd N_{\bar{\nu}}}{\dd t} = \sum_{Z,A,m} \Gamma(Z,A,m) b_{\beta}(Z,A,m) Y(Z,A,m;t) .
\end{equation}
Here $\Gamma$ is the same as in eq.~\ref{eq:time_dep_yi}, but $b_{\beta}$ is the branching ratio with which the $(Z,A,m)$ species specifically $\beta$-decays.  
We plot the total number of emitted antineutrinos per second per fission in the left panel of Fig.~\ref{fig:NuTime}.  Because the decay rate is larger for higher energy antineutrinos, those emitted at earlier times have a correspondingly higher average energy, and we plot the instantaneous average antineutrino energy as the fission yield distribution evolves in the right panel of Fig.~\ref{fig:NuTime}.

If we consider some window of time defined as an interval $[t_0,t_{f}]$, then we let number of antineutrinos emitted by $\beta$-decaying species be denoted as $N_{\bar{\nu}}$ and the total number of fissions undergone by a fissioning actinide in that window is denoted as $N_{f}$.  Then the ratio $N_{\bar{\nu}} / N_{f}$ is just given as
\begin{equation} \label{eq:anu_per_fission}
    \frac{N_{\bar{\nu}}}{N_{f}} = 
        \frac{\int_{t_0}^{t_{f}} \frac{\dd N_{\bar{\nu}}}{\dd t} \dd t }{\int_{t_0}^{t_{f}} \frac{\dd N_{f}}{\dd t} \dd t } \, .
\end{equation}
While a somewhat simple restatement, casting the ratio of antineutrinos-per-fission in this form allows us to more clearly elucidate the difference in this ratio for pulsed vs steady-state events.
When the number of fissions varies in the time window, one generally must know in detail both how the fission product yields vary as a function of time and how the fissions vary in time.  For a fission pulse at $t_{0}$, 
$\frac{\dd N_{f}}{\dd t} = N_{f} \delta(t-t_{0})$ and the neutrino emission rate will have a nontrivial time dependence as the fission product yields start their evolution at their independent yield values at $t_{0}$ and evolve toward a stable final distribution at $t_{f}$ (assuming that $t_{f}$ is chosen to be sufficiently far into the future).  Thus, generally the total number of antineutrinos per fission that one expects to be emitted from some fission environment is a nontrivial function of that environment's initial condition, subsequent evolution, and the specific window of time over which one considers.

The expected number of antineutrinos from a given reactor with a known fuel composition can be calculated from the cumulative yields of the reactor fuel components and is found to be $\approx 5$ for a typical Uranium power generation reactor.  The number of antineutrinos per fission emitted in the first $10^{3}$ seconds after the prompt fission case we find to be substantially lower (see Table~\ref{tab:pu_nu_norms_time}), however the spectrally weighted, and time integrated, cross section ($\bar{\sigma}$) is of comparable magnitude: for $^{239}$Pu fissioning in a reactor environment the Daya Bay Collaboration measured $\bar{\sigma} = 4.27 \pm 0.26 \times 10^{-43}$ cm$^{2}$ per fission \cite{DayaBay:2017jkb}, while for $^{235}$U they measured $\bar{\sigma} = 6.17 \pm 0.17 \times 10^{-43}$ cm$^{2}$ per fission to be compared with Table~\ref{tab:pu_specAvgCrossSec_time}.  A reactor fissioning in steady-state emits antineutrinos at an approximately constant rate.  The majority of these in any arbitrary window of time are of energies below the IBD threshold and so do not produce a signal in an IBD detector.  For the pulsed case, most of the neutrinos below the IBD threshold are emitted at late time, and this results in the observation that the spectrally averaged and time integrated cross sections we present in Table~\ref{tab:pu_specAvgCrossSec_time} are of a comparable values to those of a steady state reactor.

\subsubsection{Systematic Errors in Nuclear Explosive Yield}

The neutrino measurement error depends on  the number of antineutrino events detected (statistical), and systematic errors coming from  detection threshold energy resolution and reconstruction efficiency (systematic).  For statistics in the thousands of events and reasonable  assumptions on systematic  errors, we can expect a total error on the measured antineutrino flux of about 4\%, or less.  

A more uncertain error comes from extrapolating the measured antineutrino flux back to a test device explosive yield.   A few of the assumptions were discussed above.  These include errors from the prompt fission yield assumption, the detailed evolution of the neutron flux that is inducing fission, uncertainties in evaluated nuclear data and reaction libraries, and fission product database uncertainties (ENDF~\cite{Chadwick20112887short, capote2009ripl}).  These assumptions need to be tested and measured in a controlled experimental setup as described in the next section.

\section{Near-term Pulsed Reactor Demonstrator Experiment}

In the absence of nuclear weapons testing for the foreseeable future, many of the technical and theoretical challenges of a full nuclear test could be mitigated in the next few years with a smaller scale moderate cost demonstration detector.  The plan is to deploy a 20\,ton fiducial mass (25\,tons total) IBD detector placed near a pulsed reactor, such as the Texas A\&M TRIGA 1GW-10\,millisecond pulsed facility. The short duty cycle and repeatability of pulses would provide critical real condition testing and knowledge which would be valuable for planning  a possible real test  shot in the future. 
Furthermore, the measured neutrino rate per pulse as a function of time would provide unique constraints on fission databases, equilibrium assumptions, and yield errors. Such a near-term demonstrator experiment could be prepared quickly and at a reasonable cost.

Typically, a pulsed reactor can generate $\sim10^{17}$ fissions in a single pulse, thus many pulses are needed to achieve the number of fissions required for the demonstration. By recording events triggered in a $\sim$1000 seconds timing window following a reactor pulse, backgrounds can be further reduced and a decay curve of the fission products can be measured.
The close proximity of the detector and the ability offline to superimpose multiple pulses on the same pulse trigger allows an accurate simulation of the event rate pile-up in a nuclear explosion. In Section~\ref{sec:PNS}  and \ref{pulsedbackgrounds}, we will discuss various pulsed neutrino sources that we have explored and the number of antineutrinos that we expect to see at each facility. Section~\ref{technical} details the technical measurements to be made, Section \ref{sec:Detector} will describe our preliminary detector design, and \ref{detectorsim} discusses detector response and simulations. The size of the detector has to allow for sufficient statistics based on the total number of fissions that we expect at each facility.  

\subsubsection{Pulsed Neutrino Sources}
\label{sec:PNS}
One of the pulsed reactor sources that we have identified is TRIGA reactors, which can run in steady-state and pulsed mode~\cite{Triga}.
The pulsed mode is accomplished by rapidly firing the control rod out of the core within a fraction of a second. The reactor quickly produces a super-criticality pulse from a low power baseline. As the temperature of the fuel increases, its reactivity decreases and allows the reactor to return to normal levels, the reactor is then automatically scrammed after a few seconds to clip the power tail for operational
convenience \cite{Boeck2007}. 
This brief supercritical pulse of fission can generate a tremendous antineutrino flux over 10's of seconds, which is ideal for separation from the prompt fast neutron and gamma related backgrounds and steady state random backgrounds.  

\begin{figure} [h] 
 \centering
\includegraphics[width=3.25in]{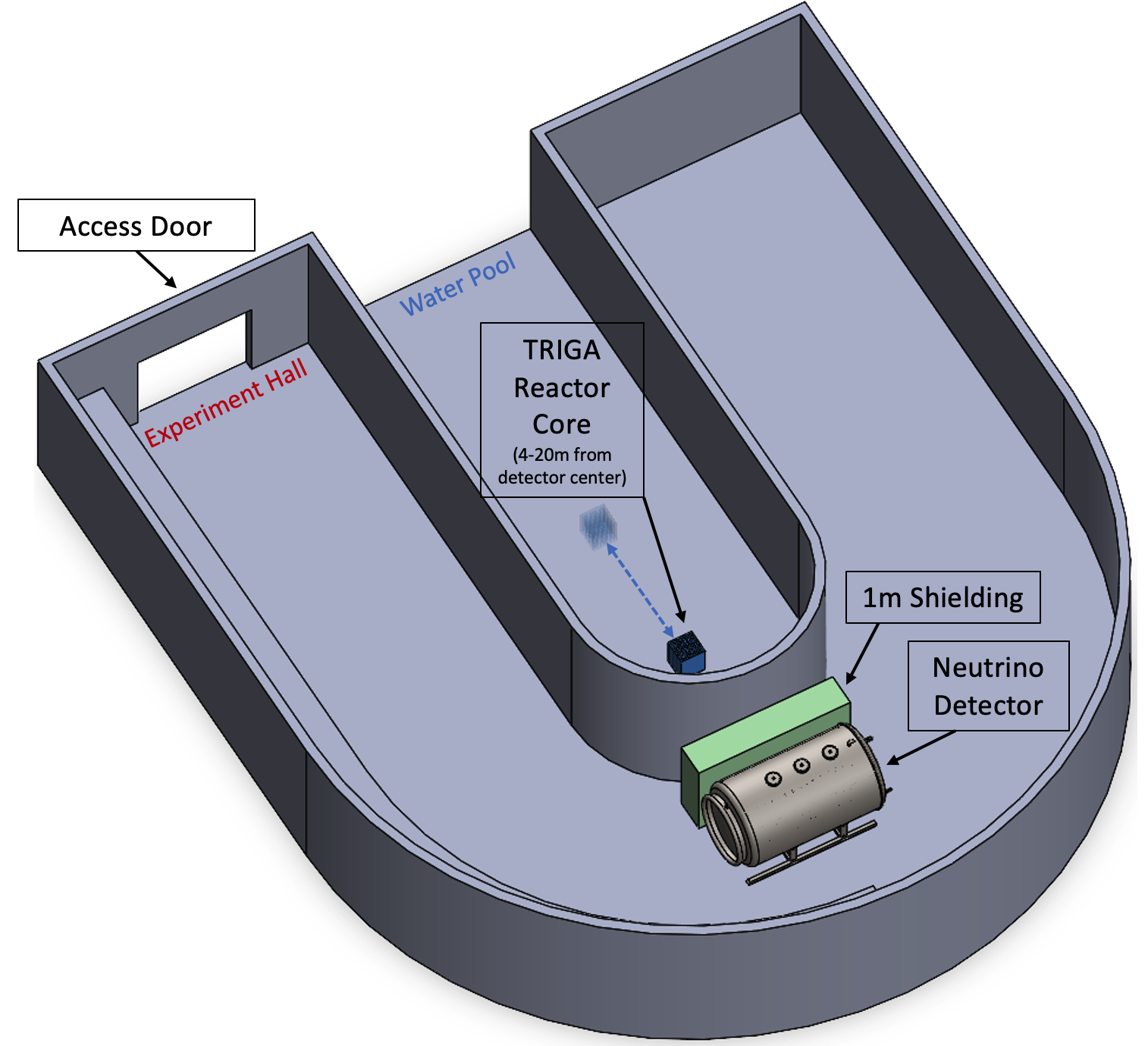}
 \caption{The 20-ton (fiducial) mass $\nu$FLASH detector situated at  the TAMU TRIGA reactor. There is sufficient space to place the detector as close as 4\,m from the movable reactor and to allow 1\,m of  shielding (combination of borated poly, concrete, and steel).  If required, the detector can be placed further away from the reactor to allow more shielding.   There is sufficient facility infrastructure (power, cooling, etc) to mount the experiment effectively. }
  \label{fig:NuFLASHatTRIGA}
\end{figure}

The Texas A\&M (TAMU) TRIGA reactor has been identified as a potential neutrino source. The 20\% $^{235}$U enriched reactor has a steady-state fission rate of 1-megawatt and can produce a peak power 250-megawatt pulse with a width of up to 40\,ms, producing $3\times10^{17}$ fissions per pulse~\cite{Triga}. The reactor also has large floor  space and experimental facilities to deploy a moderate size detector as close as 4\,m, including shielding, from the reactor core (see Figure~\ref{fig:NuFLASHatTRIGA}). The pulsed reactor at Sandia National Laboratory~\cite{Parma2019} has even higher pulsed power characteristics as TAMU, greater than $\sim 1\times10^{18}$ fissions per pulse.  We will also be investigating the logistical feasibility of mounting a neutrino detector at the Sandia pulsed reactor.

\subsubsection{IBD and Background event rates at TAMU TRIGA}
\label{pulsedbackgrounds}

The expected number of  fissions per pulse at the TAMU TRIGA during normal operations is $3\times10^{17}$ fissions/pulse.   Due to small neutrino rates per pulse, analysis will only take place between 0.1 and 100 seconds after a pulse to reduce prompt and steady state backgrounds (the DAQ window will still be active for at least 1000 seconds).  Calculating $\bar{\sigma}$, from Table~\ref{tab:pu_specAvgCrossSec_time}, for a prompt fissioning of $^{235}$U from thermal neutrons in this time range  gives $5.32 \times 10^{-43}$ cm$^2$.  
Given a detector distance of 4\,m (center of reactor to center of detector), and the same number of free $H$ per detector ton from Section~\ref{sec:Estimate}, then the rate is 0.0068 antineutrinos/ton/pulse (this already includes threshold efficiencies).   
For 2000 pulses per operational year~\cite{TAMUMAnagement}, a 20-ton fiducial detector discussed below, and 80\% detection efficiency (Section~\ref{sec:Estimate}), then the IBD event rate is 218 events/year.  This is a small but reasonable amount of events to test the various technical challenges of detecting neutrinos from a short burst.  
A three year run will provide a 4\% statistical error on the measured pulsed power yield. Given the higher power, the event rate at Sandia could be much larger if the detector can be placed at a similar nearby distance as at TAMU. 


Shielding will be added to reduce both fast and thermal neutrons and gamma-rays from the reactor pulse.  However, it is still anticipated that a significant number will make it to the detector at such a close distance.  The strategy to mitigate this prompt component is to cut out from analysis the first 100\,milliseconds of data.  This will result in only a 2.7\% reduction in antineutrino signal since the majority come from slower fission decays (see Table~\ref{tab:pu_specAvgCrossSec_time}).

Accidental backgrounds from cosmic rays (500\,Hz) and PMT beta-gamma's (2\,kHz) are easily rejected by external muon paddle detectors and PMT large charge cuts and will create an analysis dead-time of only 0.25\%.  Steady state background from activation and other external source of $\sim$MeV gamma-rays is expected to be at most 10\,kHz (based on the 10-ton LAr CCM detector  measurements at Lujan neutron source~\cite{CCM:2021leg}). Adding appropriate shielding including borated poly (for neutron capture) and steel (for gamma absorption) should reduce this to $\sim$ 3.5\,kHz, again from CCM experience.
After the 1.8\,MeV energy threshold and fiducial cuts this can be reduced to $<$100\,Hz. Finally, accidental  neutrons from cosmogenic sources~\cite{https://doi.org/10.1029/2012JA017524} (mimic IBD via neutron capture coincidence with random gamma-ray or neutron scatter producing gamma-ray then thermalizing and capture) and PMT glass~\cite{ZHANG201867} sources are estimated to be $\sim$Hz rate.
At this level the accidental background rates to IBD (for a 10 microsecond neutron capture time) from external backgrounds is 0.1\%.  Using a 100 second DAQ window pulsed 2000 times gives 200 events in a one year run.  As this is steady state it can be measured precisely by strobe  and pre-pulse data and subtracted from signal and represents a $\pm 14$ event statistical subtraction error, or about $\sim7\%$ on the expected signal.  However, adding more shielding and improved analysis  to control the rate is important to reducing the statistical error in the subtraction.

\subsubsection{IBD Technical Tests Measurements at TAMU TRIGA}
\label{technical}

The primary purpose of running an IBD detector at  a pulsed reactor is to test and understand the characteristics and issues related to IBD neutrino detection from a short pulsed source.  In the past all IBD neutrino detector experiments have been performed at steady state reactors.  The capability of these detectors have been well characterized and proven.  However, a pulsed source such as an nuclear weapons test has some unique challenges.  The following is a list of tests to be performed for an IBD detector at a pulsed reactor source,

\begin{itemize}
\item Most of the antineutrinos arrive within 1000 seconds.  The preferred run mode is an unbiased trigger (collecting all digitization data for all PMT's over the entire 1000 seconds) which puts an extremely large data throughput load on the DAQ ($>$1000\,TBytes/pulse).  The heavy data throughput will require new solutions such as faster readout hardware, more DAQ computers, and an increased fast data storage buffer that  all need to be tested and implemented.  If an unbiased trigger cannot be implemented, then an event trigger based on  detector  activity will have to be  studied and implemented.  There can be a loss of physics if this is not done carefully and tested before a potential one-time shot.

 \item  Large pile up of IBD events from a weapons test up to a 1000 second interval will present reconstruction challenges.  An IBD event is identified by a prompt positron then followed by a capture neutron on Gd within about 10\,microseconds.   With multiple events happening together (up to many thousands from a nuclear explosion), uniquely tagging a positron with a correlated capture neutron becomes more difficult.  At a low rate steady state reactor this is not an issue, but for an intense  pulsed source it becomes critical.   Any one pulse from the TRIGA reactor will have minimal event overlap, but after collecting a few years worth of  many pulses, these events can be overlaid to produce a data stream that mimics the higher intensity with a bomb.  Analysis strategies can then be developed and tested before a one-time shot.
A potential outcome of this study is the realization that capturing the IBD neutron with Gd will only confuse the event reconstruction and is best not added to the scintillator.
\item Perform background measurements of both steady state (cosmic rays, cosmogenic neutrons, PMT U/Th activity, neutron activation, etc) and direct backgrounds from the source such as fast neutrons and gamma-rays. Develop strategies to mitigate.


\item Starting  with the IBD antineutrino rate measurement, we will apply reactor fission codes (such as ENDF~\cite{Chadwick20112887short}) to extrapolate back to the reactor power per pulse, which is known to $\sim$1\%  level~\cite{TAMUMAnagement}.  This is important for testing the various methods and codes necessary for bomb test yield predictions using antineutrino rates.  From steady state reactor measurements in the past we know this error to 3-4\%.  
The TAMU TRIGA reactor runs half the time in steady state mode which will allow a measurement of the steady state antineutrino rate (this is estimated to be over 10,000 events/year, albeit with much higher background rate).  Comparison of the steady state rate to pulsed mode with the same reactor and detector could yield useful information on equilibrium fission rates versus pulsed dynamics.

\end{itemize}

There is also a set of basic science measurements that can be performed parasitically to the main effort and is described in Section~\ref{science}.

\subsubsection{IBD Test Detector Design ($\nu$FLASH)}
\label{sec:Detector}

To achieve the estimated 654 IBD antineutrino event interactions in three years of pulsed running at the TAMU TRIGA reactor requires at least a 20-ton fiducal mass (see calculations above).  Due to mass losses in the outer veto regions and lack of efficiency, a detector with 25-tons total mass is required.  The detector is to be filled with the commercially available EJ-335 scintillator oil doped with Gd~\cite{eljen335}.  
This oil has proven high scintillation light output of 9600 photons/MeV of deposited energy and a thermal neutron capture time of 5.5 microseconds for a Gd concentration of 0.5\% ($\sim$30\,microseconds for 0.1\% Gd concentration). The higher the Gd concentration the shorter the capture time and better background rejection from the positron-neutron coincidence.  

\begin{figure}[h] 
   \centering
\includegraphics[width=3.5in, angle= 90]{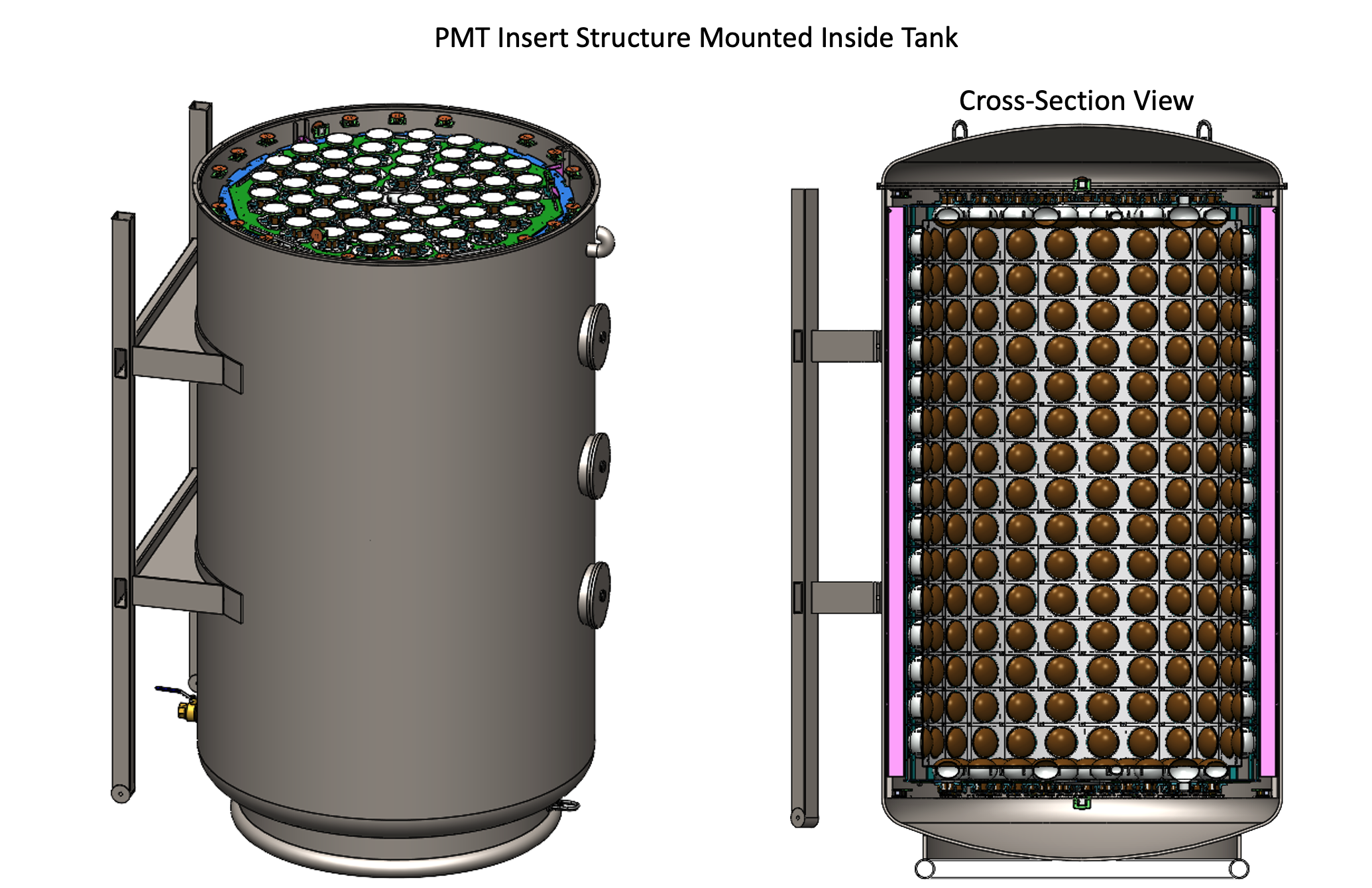} 
   \caption{The pulsed reactor $\nu$FLASH test detector design.  A total of 25-ton mass (8.5' diameter and 20' length) of gadolinium doped mineral oil (EJ-335) and instrumented with up to 500  8" PMTs. The fiducial volume for efficient neutrino detection is 20-tons. Most of the physical components have been proven with CCM and detection methods demonstrated by other reactor neutrino experiments. }
   \label{fig:NuFLASH}
\end{figure}

The detector physical design (see Figure~\ref{fig:NuFLASH}) is based  on the successful Coherent  Captain-Mills (CCM) experiment which is performing accelerator-based sterile neutrino, dark matter, and axion searches at LANSCE Lujan center~\cite{CCM:2021leg,CCM:2021yzc,CCM:2021jmk}.   Even though  the CCM experiment is liquid argon  based, the  physical structures, PMT's, electronics, data analysis, and simulations will share  many of the physics features of the $\nu$FLASH detector.   This will save an enormous amount of design time and experimental effort based on a proven sensitive detector concept. Figure~\ref{fig:NuFLASH} shows a solidworks design for the detector  that is 20' long  by 8.5' diameter (25-tons oil).  The diameter is chosen to be transportable on a flat bed truck so it can be conveniently built  at LANL then sent to TAMU, or elsewhere should future sources be envisioned. 

We plan to use up to 500 standard 8" diameter PMT's which have 22\% quantum efficiency around 400\,nm to detect the scintillation light produced by the IBD final state particle interactions (position and neutron capture gamma-rays). These will be instrumented with CAEN V1730 (500 MHz 14 bit  charge ADCs) which will measure  the antineutrino interaction light pulses to nanosecond time  resolution and better than 5\% charge resolution.

\subsubsection{Detector Simulation Results}\label{detectorsim}

Preliminary simulations of the $\nu$FLASH mineral oil detector have been completed to show the feasibility of this method. Initial simulations used the existing CCM geometry~\cite{thesis_Edward,CCM:2021jmk} and Geant4~\cite{geant4} framework to test the ability of such a detector to identify IBD coincidence between the initial positron and the successive neutron capture. 

The simulation used Gd-doped EJ-335 mineral oil at various concentrations. The mineral oil filled a cylindrical detector with 200 internal PMTs facing a 7-ton fiducial volume. An incident antineutrino spectrum was convoluted with the IBD capture cross section to produce an event spectrum. These events consisted of $n + e^+$ generated simultaneously with appropriate energy/momentum from a antineutrino IBD event. Both initial particles were then allowed to propagate using the FTFP\_BERT\_HP physics list and Geant4 optical physics built from the properties of EJ-335 mineral oil scintillator~\cite{eljen335}. Detection occurred through optical photon interaction with models of the current CCM PMTs built from LED-based single photo-electron (SPE) and radioactive source-based efficiency calibrations~\cite{thesis_Edward,CCM:2021jmk}. 

\begin{figure} [h] 
 \centering
\includegraphics[width=3.5in]{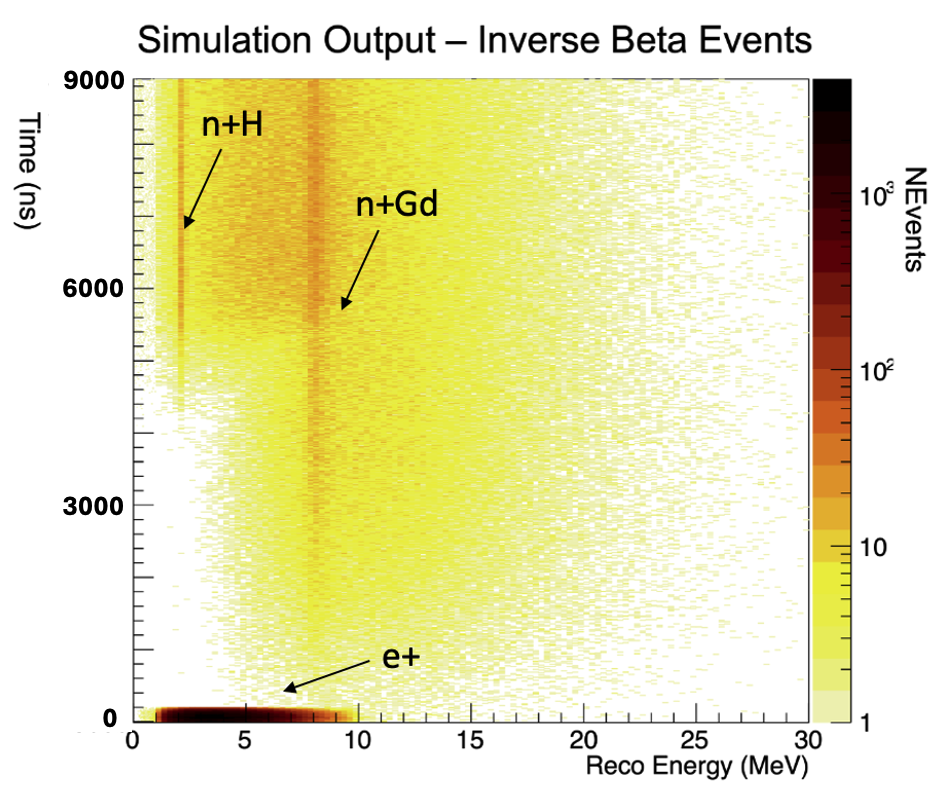}
 \caption{The reconstructed time and energy output of the IBD simulation in the $\nu$FLASH detector.  At the bottom left, in dark red, is the fast positron scintillation peaking at $\sim$2.4 MeV. The large cloud that dominates the figure is the delayed neutron captures out to 9 us after the positron. The majority of such captures are Gd-captures producing 8 MeV of gammas. A minority are hydrogen captures producing 2.2 MeV, producing a second line above the positron cluster.}
  \label{fig:ibSim2d}
\end{figure}

The simulation produced expected results - see Figure \ref{fig:ibSim2d}. 100\% of initial positrons were detected to some level, although a small fraction partially escaped the fiducial volume adding to the energy smearing. The positron spectrum peaked at $\sim$3000 PEs (photo-electrons, a measure of reconstructed energy), indicating a true/reco energy ratio of 1 MeV to 1250 PEs. This gives the simulated detector sensitivity to energies between 0.01 and 30 MeV before PMT non-linearity and saturation may become an issue. 
We expected a $\sim5$ $\mu$s neutron thermalization time and a $\sim6$ $\mu$s thermal neutron capture time for 1\% Gd doping from similar processes in water \cite{FUJINO1970,Dumazert2018}, with a longer capture time for lower concentrations of Gd. 
The simulation demonstrated this behavior.  Approximately 80\% of neutrons captured within a 10 $\mu$s window at 1\% Gd doping, while using 0.5\% doping only captured $\sim$35\% within the same window.   The majority of captures were on Gd-155 or Gd-157 in both cases, producing a cascade of gamma-rays totaling approximately $\sim$8 MeV. 
This was reconstructed in the expected energy region compared to the positron spectrum, between 8,000 and 10,000 PEs, with the spread coming from $\sim$20\% energy resolution smearing. This high uncertainty is derived from the multi-particle cascade, sometimes producing much higher reconstructed energy up to 35000 PE $\approx$ 30 MeV.  A fraction of the captures were on H, producing a second feature in Figure \ref{fig:ibSim2d} at $\sim2$ MeV and 6 $\mu$s. The positron's reconstructed time is taken as the antineutrino arrival time since the positron is produced immediately after interaction. 
The antineutrino  energy is taken as the reconstructed positron energy plus 1.8\,MeV due to interaction threshold.  The energy resolution  on  the reconstructed positron is about 5\%  within the fiducial volume.  This can be  further reduced with advanced reconstruction  techniques such as charge-time log-likelihood reconstruction analysis.  For the antineutrino rate calculations in this paper we assume 1\% Gd doping giving an overall IBD detection efficiency of 80\%.

\subsubsection{Additional Fundamental Science Drivers}
\label{science}

\begin{figure}[htbp] 
 \vspace{-1ex}
   \centering
   \includegraphics[width=3.0in]{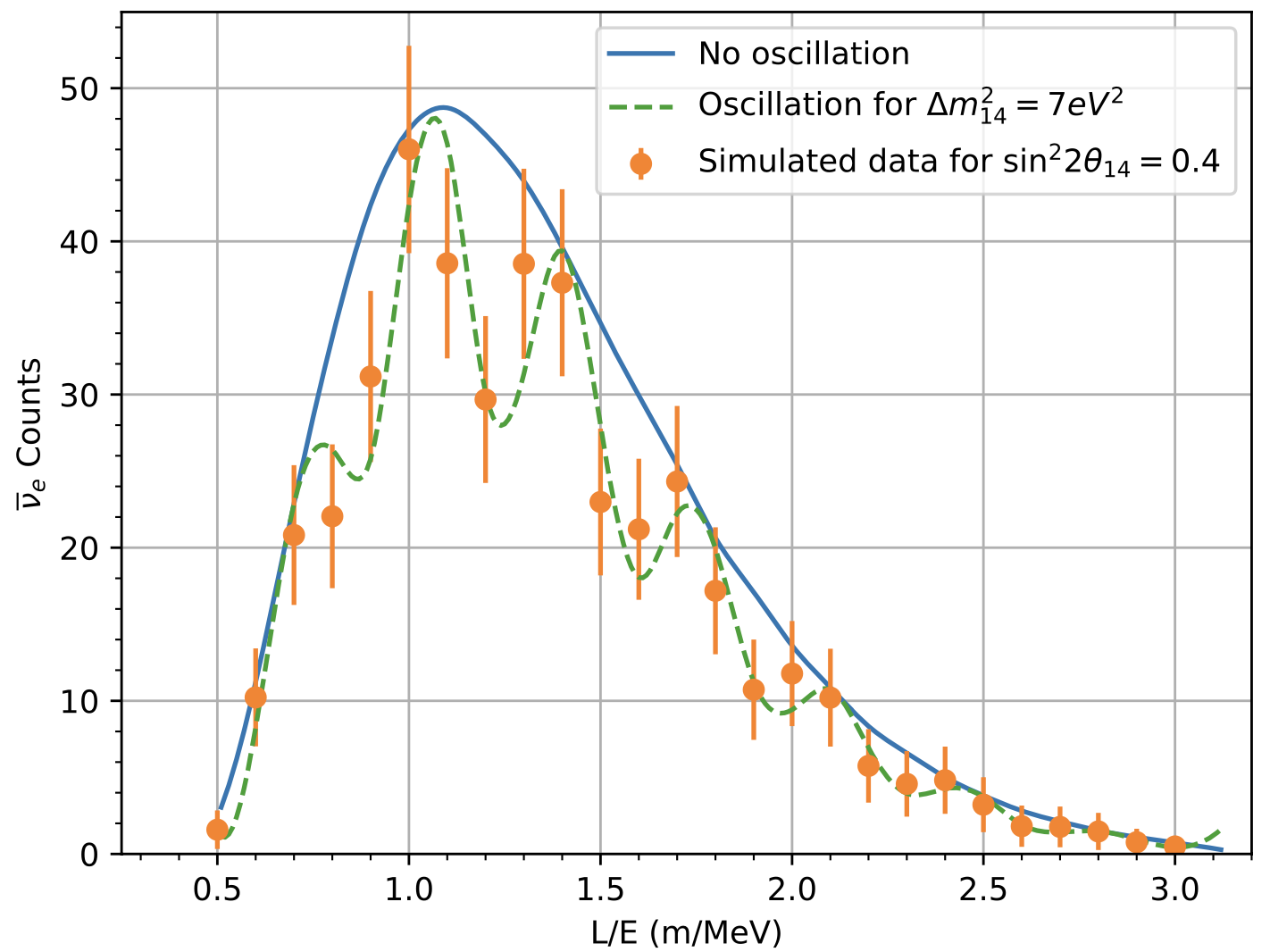} 
    \includegraphics[width=2.3in]{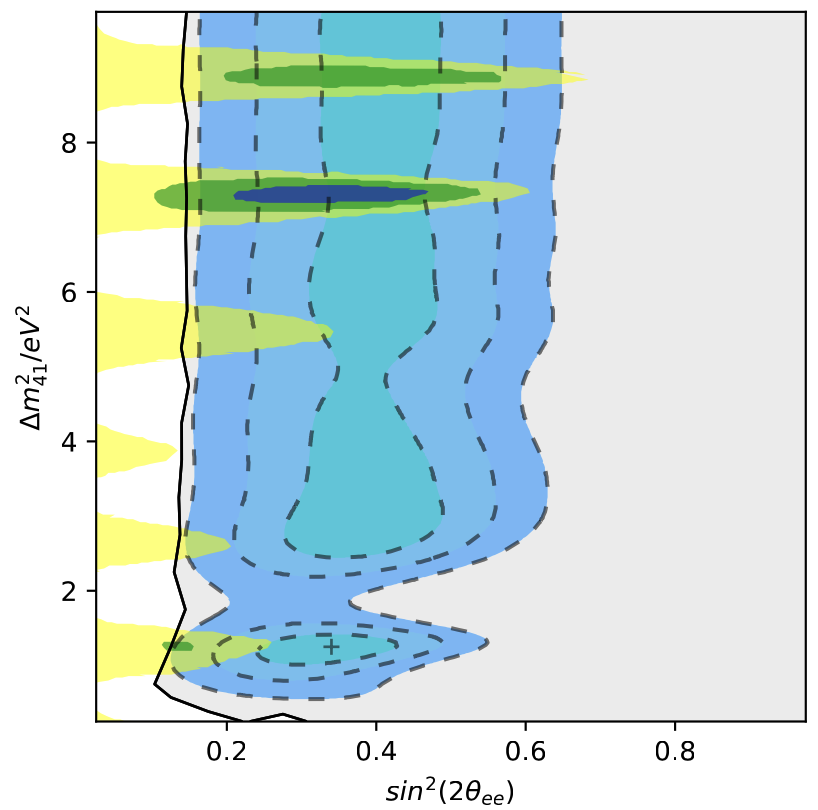} 
    \vspace{-1.25ex}
   \caption{Left: Simulated  $\nu$FLASH 20-ton IDB detector neutrino L/E distribution after three years of running at the TAMU pulsed reactor at a distance of 4\,m from the source.  Error bars are statistical only, and assumes $\Delta m^2_{14}= 7 eV^2$ and $\sin^2 2\theta_{14}= 0.4$ (near Neutrino-4 best fit).    Right: Three year 90\% confidence limits sensitivity from simultaneous rate normalization and energy shape fit for $\nu$FLASH~(solid line and grey region) covering the signal regions from Neutrino-4 (Yellow) and BEST (Blue). }
   \vspace{0ex}
   \label{fig:SENS}
\end{figure}

Parasitic to testing technical capabilities of an IBD detector at the TAMU TRIGA pulsed source for weapons physics, there are a number of  basic science measurements that can be pursued in parallel. This has the added benefit of attracting university collaborators and students to participated in the open science and help make the overall project a success.  The not  complete list of science measurements are,
\begin{itemize}
\item Investigate the source of the $\sim$5 MeV bump in the reactor antineutrino spectrum observed at steady state reactor experiments~\cite{Huber:2016xis}.  
One interesting measurement is the time profile might yield information on a prompt new BSM process that can be separated in time from slow fission antineutrino sources. However, statistics will be a challenge for $\nu$FLASH since the bump is only $\sim$10\% of the total neutrino rate and is more a venue for future research.
\item The energy and baseline of the $\nu$FLASH experiment is ideal to test for sterile neutrino oscillations at  $L/E \sim 1$\,m/MeV region~\cite{Vogel:2015wua}.   This could provide a direct measurement of  neutrino oscillations consistent with Neutrino-4 and BEST results~\cite{Serebrov:2020kmd,Barinov:2022wfh}  that  can be interpreted as $\bar{\nu}_e$ or $\nu_e$ disappearance.  
With a planned three year run, Figure~\ref{fig:SENS} shows a toy model anticipated reach of $\nu$FLASH to search for large $\Delta m^2_{14}$ $\overline{\nu}_e$ disappearance oscillations consistent with the Neutrino-4 and BEST signals.  With 2000 pulses per year and detector at 4\,m, we anticipate 218 unoscillated IBD antineutrinos/year after efficiency as discussed previously in Section~\ref{pulsedbackgrounds}. Due to the compact core,  $\nu$FLASH can perform a sensitive L/E shape analysis where many of the large systematic errors from antineutrino flux and cross section only affect normalization, not shape, to first order. Also, due to the short source pulsing and IBD signal coincidence, backgrounds will be mitigated significantly and are assumed to be small - the sensitivity plot in Fig~\ref{fig:SENS} includes 10\% background subtraction error (see subsection~\ref{pulsedbackgrounds} for details).  Other errors included is the position smearing effects from source size of 0.5\,m on each side (standard deviation of 0.15\,m) and event position reconstruction of 10\,cm position resolution.  Finally, the IBD spectrum energy is smeared by 5\% to account for the detector energy reconstruction resolution. If an oscillation signal is observed, then the movable reactor core position could be changed to optimize sensitivity at the best fit point.
Further details of the experiment and sensitivities will be presented in a future physics paper.  The irony here is if sterile neutrinos at the $\Delta m^2_{14} \sim 1 eV^2$ do exist, they would become the largest systematic error in nuclear weapons yield extrapolation from neutrino rate measurements. 

\item A pulsed reactor, and nuclear weapon blast, produces a large number of prompt (less than a few hundred nanoseconds) gamma-rays and electrons~\cite{Gatera:2018tim} that can potentially couple to axions or axion like particles (ALP).  Recent models have demonstrated the $\sim$1 MeV region as a motivated place to search~\cite{Dent:2019ueq}.   
The prompt time production of axions/ALPs in the reactor core will help separate it from the IBD antineutrinos coming from the slower fission decays, but will still have to contend with large prompt backgrounds from the beam neutrons and gamma-rays.  The limited pulsing of the reactor will reduce signal rate, but the significant random background rejection due to the pulse timing will improve overall sensitivity.   Sensitivity calculations are on-going and will be reported in a future paper.

\end{itemize}

\section{Conclusions}

  Due to the prolific production of neutrinos in fission reactions during a nuclear weapons test, they can be used as a novel and complementary tool for diagnostics.  With large fiducial mass and prolific scintillation light detection, modern neutrino detector using the IBD reaction can easily detect thousands of antineutrinos from a stand off distance of 500 meters.  Should the U.S. ever go back to nuclear weapons testing, the use of neutrinos would provide a new tool to better understand and diagnose the performance of a weapon.   In the meantime, the technology and methods used can be tested at a pulsed reactor such as the TAMU TRIGA reactor, which is a good surrogate for a nuclear weapons test.  This enables studying yield extraction, non-equilibrium fission rates, and systematic errors. As well, tests can be performed of the IBD neutrino detector performance in a pulsed environment, such as neutrino pileup, readout rates, triggering, reconstruction, etc.  Finally, with a modest sized 20-ton fiducial mass scintillation detector ($\nu$FLASH) deployed at a pulsed reactor, tests of the Neutrino-4 and BEST disappearance signal can be performed, which could be indication of sterile neutrino oscillations at the $\sim 1eV^2$ scale.  Using a reactor in pulsed mode for neutrino physics is novel and would yield significant background reduction and enable the study of the time evolution of fission neutrinos. Such a rich program is of modest cost and could be deployed in a few years.

\section{Acknowledgements}

We thank Prof.\ B.~Dutta (TAMU) and I.~Bisset (TAMU) for physics discussions and generating the sterile neutrino sensitivity plots in Figure~\ref{fig:SENS}.    This document has been reviewed and has LANL library reference number LA-UR-24-29431.

\vspace{0.75cm}
\bibliography{main}

\begin{thebibliography}{10}

\bibitem{Reines1953}
F.~Reines and C.~L. Cowan.
\newblock A proposed experiment to detect the free neutrino.
\newblock {\em Physical Review}, 90:492, 5 1953.

\bibitem{LANL1997}
Los Alamos.
\newblock The reines-cowan experiments: Detecting the poltergeist.
\newblock {\em Los Alamos Science}, 25:3, 1997.

\bibitem{Cowan1956}
C.~L. Cowan, F.~Reines, F.~B. Harrison, H.~W. Kruse, and A.~D. McGuire.
\newblock Detection of the free neutrino: a confirmation.
\newblock {\em Science}, 124:103--4, 7 1956.

\bibitem{ParticleDataGroup:2020ssz}
P.~A. Zyla et~al.
\newblock {Review of Particle Physics}.
\newblock {\em PTEP}, 2020(8):083C01, 2020.

\bibitem{Abbott2021}
Alison Abbott.
\newblock The singing neutrino nobel laureate who nearly bombed nevada.
\newblock {\em Nature}, 593:334--335, 5 2021.

\bibitem{Qian:2018wid}
Xin Qian and Jen-Chieh Peng.
\newblock {Physics with Reactor Neutrinos}.
\newblock {\em Rept. Prog. Phys.}, 82(3):036201, 2019.

\bibitem{Note1}
Throughout this work we use the convention kTe for kilo-tons of explosive
  yield, while for neutrino detector mass we use ton.

\bibitem{Huber:2011wv}
Patrick Huber.
\newblock {On the determination of anti-neutrino spectra from nuclear
  reactors}.
\newblock {\em Phys. Rev. C}, 84:024617, 2011.
\newblock [Erratum: Phys.Rev.C 85, 029901 (2012)].

\bibitem{Marti:2019dof}
Ll. Marti et~al.
\newblock {Evaluation of gadolinium\textquoteright{}s action on water Cherenkov
  detector systems with EGADS}.
\newblock {\em Nucl. Instrum. Meth. A}, 959:163549, 2020.

\bibitem{DayaBay:2015kir}
F.~P. An et~al.
\newblock {The Detector System of The Daya Bay Reactor Neutrino Experiment}.
\newblock {\em Nucl. Instrum. Meth. A}, 811:133--161, 2016.

\bibitem{DoubleChooz:2022ukr}
H.~de~Kerret et~al.
\newblock {The Double Chooz antineutrino detectors}.
\newblock {\em Eur. Phys. J. C}, 82(9):804, 2022.

\bibitem{Anna}
Anna Hayes and Josh Martin~(LANL T-2).
\newblock {\em Simulation estimate.}

\bibitem{krane1991introductory}
Kenneth~S Krane.
\newblock {\em Introductory nuclear physics}.
\newblock John Wiley \& Sons, 1991.

\bibitem{osti_1340968}
Susan~Kloek Hanson, Anthony~Douglas Pollington, Christopher~Russell Waidmann,
  William~Scott Kinman, Allison~Marie Wende, Jeffrey~L. Miller, Jennifer~A.
  Berger, Warren~James Oldham, and Hugh~D. Selby.
\newblock Measurements of extinct fission products in nuclear bomb debris:
  Determination of the yield of the trinity nuclear test 70 y later.
\newblock {\em Proceedings of the National Academy of Sciences of the United
  States of America}, 113(29), 7 2016.

\bibitem{an2017improved}
Feng~Peng An, Akif~Baha Balantekin, Henry~Reyer Band, M~Bishai, Simon Blyth,
  D~Cao, Guo~Fu Cao, Jun Cao, WR~Cen, Yat~Long Chan, et~al.
\newblock Improved measurement of the reactor antineutrino flux and spectrum at
  daya bay.
\newblock {\em Chinese Physics C}, 41(1):013002, 2017.

\bibitem{okumura2022energy}
Shin Okumura, Toshihiko Kawano, Amy~Elizabeth Lovell, and Tadashi Yoshida.
\newblock Energy dependent calculations of fission product, prompt, and delayed
  neutron yields for neutron induced fission on 235u, 238u, and 239pu.
\newblock {\em Journal of Nuclear Science and Technology}, 59(1):96--109, 2022.

\bibitem{capote2009ripl}
Roberto Capote, Michel Herman, P~Oblo{\v{z}}insk{\`y}, PG~Young, St{\'e}phane
  Goriely, T~Belgya, AV~Ignatyuk, Arjan~J Koning, St{\'e}phane Hilaire,
  Vladimir~A Plujko, et~al.
\newblock Ripl--reference input parameter library for calculation of nuclear
  reactions and nuclear data evaluations.
\newblock {\em Nuclear Data Sheets}, 110(12):3107--3214, 2009.

\bibitem{Chadwick20112887short}
M.B. Chadwick, M.~Herman, P.~Oblo{\v z}insk{\' y}, et~al.
\newblock {ENDF/B-VII.1} nuclear data for science and technology: Cross
  sections, covariances, fission product yields and decay data.
\newblock {\em Nuclear Data Sheets}, 112(12):2887 -- 2996, 2011.
\newblock Special Issue on ENDF/B-VII.1 Library.

\bibitem{HayesVogel}
Anna~C. Hayes and Petr Vogel.
\newblock Reactor neutrino spectra.
\newblock {\em Annual Review of Nuclear and Particle Science}, 66(1):219--244,
  2016.

\bibitem{Note2}
We have further assumed that all considered beta decays are allowed. A more
  detailed analysis including shape factor corrections to forbidden decays will
  be presented in a future work.

\bibitem{DayaBay:2017jkb}
F.~P. An et~al.
\newblock {Evolution of the Reactor Antineutrino Flux and Spectrum at Daya
  Bay}.
\newblock {\em Phys. Rev. Lett.}, 118(25):251801, 2017.

\bibitem{Triga}
{TRIGA Reactor Characteristics},
  https://ansn.iaea.org/Common/documents/Training/TRIGA

\bibitem{Boeck2007}
H.~Boeck, M.~Villa, and Vienna (Austria) Vienna University of~Technology Atomic
  Institute of~the Austrian~Universities.
\newblock Triga reactor characteristics, 1 2007.

\bibitem{Parma2019}
Edward J.~Parma Jr. and Michael~Warren Gregson.
\newblock {The Annular Core Research Reactor (ACRR) Description and
  Capabilities}.
\newblock {https://www.osti.gov/servlets/purl/1761930 (2019)}.

\bibitem{TAMUMAnagement}
TAMU~TRIGA Management.
\newblock {\em Communicated estimate based on pulsed running in the evenings}.

\bibitem{CCM:2021leg}
A.~A. Aguilar-Arevalo et~al.
\newblock {First dark matter search results from Coherent CAPTAIN-Mills}.
\newblock {\em Phys. Rev. D}, 106(1):012001, 2022.

\bibitem{https://doi.org/10.1029/2012JA017524}
W.~Rühm, U.~Ackermann, C.~Pioch, and V.~Mares.
\newblock Spectral neutron flux oscillations of cosmic radiation on the earth's
  surface.
\newblock {\em Journal of Geophysical Research: Space Physics}, 117(A8), 2012.

\bibitem{ZHANG201867}
Xuantong Zhang, Jie Zhao, Shulin Liu, Shunli Niu, Xiaoming Han, Liangjian Wen,
  Jincheng He, and Tao Hu.
\newblock Study on the large area mcp-pmt glass radioactivity reduction.
\newblock {\em Nuclear Instruments and Methods in Physics Research Section A:
  Accelerators, Spectrometers, Detectors and Associated Equipment}, 898:67--71,
  2018.

\bibitem{eljen335}
Eljen~Technology (EJ-335),
  https://eljentechnology.com/products/liquid-scintillators/ej-331-ej-335.

\bibitem{CCM:2021yzc}
A.~A. Aguilar-Arevalo et~al.
\newblock {First Leptophobic Dark Matter Search from the
  Coherent\textendash{}CAPTAIN-Mills Liquid Argon Detector}.
\newblock {\em Phys. Rev. Lett.}, 129(2):021801, 2022.

\bibitem{CCM:2021jmk}
A.~A. Aguilar-Arevalo et~al.
\newblock {Prospects for detecting axionlike particles at the Coherent
  CAPTAIN-Mills experiment}.
\newblock {\em Phys. Rev. D}, 107(9):095036, 2023.

\bibitem{thesis_Edward}
{Dunton, Edward C.}
\newblock A search for axion-like particles at the coherent captain mills
  experiment (https://doi.org/10.7916/x9x1-ka48).
\newblock 2022.

\bibitem{geant4}
Geant4 Physics~Simulation Code.
\newblock {\em NIM A 506 (2003), pg 250.}

\bibitem{FUJINO1970}
Michihira FUJINO and Kenji~SUMITA and.
\newblock Measurements of neutron thermalization time constant of light water
  by pulsed neutron method.
\newblock {\em Journal of Nuclear Science and Technology}, 7(6):277--284, 1970.

\bibitem{Dumazert2018}
Quentin Lecomte Guillaume H.V. Bertrand Matthieu~Hamel Jonathan~Dumazert,
  Romain~Coulon.
\newblock Gadolinium for neutron detection in current nuclear instrumentation
  research: A review.
\newblock {\em Nuclear Instruments and Methods in Physics Research Section A:
  Accelerators, Spectrometers, Detectors and Associated Equipment}, 882:53--68,
  2018.

\bibitem{Huber:2016xis}
Patrick Huber.
\newblock {NEOS Data and the Origin of the 5 MeV Bump in the Reactor
  Antineutrino Spectrum}.
\newblock {\em Phys. Rev. Lett.}, 118(4):042502, 2017.

\bibitem{Vogel:2015wua}
Petr Vogel, Liangjian Wen, and Chao Zhang.
\newblock {Neutrino Oscillation Studies with Reactors}.
\newblock {\em Nature Commun.}, 6:6935, 2015.

\bibitem{Serebrov:2020kmd}
A.~P. Serebrov et~al.
\newblock {Search for sterile neutrinos with the Neutrino-4 experiment and
  measurement results}.
\newblock {\em Phys. Rev. D}, 104(3):032003, 2021.

\bibitem{Barinov:2022wfh}
V.~V. Barinov et~al.
\newblock {Search for electron-neutrino transitions to sterile states in the
  BEST experiment}.
\newblock {\em Phys. Rev. C}, 105(6):065502, 2022.

\bibitem{Gatera:2018tim}
Ang\'elique Gatera, Alf G\"o\"ok, Franz-Josef Hambsch, Andr\'e Moens, Andreas
  Oberstedt, Stephan Oberstedt, Goedele Sibbens, David Vanleeuw, and Marzio
  Vidali.
\newblock {Prompt fission\ensuremath{\gamma}-ray characteristics from
  neutron-induced fission on 239Pu and the time-dependence of
  prompt-\ensuremath{\gamma}ray emission}.
\newblock {\em EPJ Web Conf.}, 169:00003, 2018.

\bibitem{Dent:2019ueq}
James~B. Dent, Bhaskar Dutta, Doojin Kim, Shu Liao, Rupak Mahapatra, Kuver
  Sinha, and Adrian Thompson.
\newblock {New Directions for Axion Searches via Scattering at Reactor Neutrino
  Experiments}.
\newblock {\em Phys. Rev. Lett.}, 124(21):211804, 2020.

\end{thebibliography}
\bibliographystyle{unsrt}

\end{document}